\documentstyle[12pt]{article}
\def\gsim{\mathrel{%
\rlap{\raise 0.511ex \hbox{$>$}}{\lower 0.511ex
\hbox{$\sim$}}}}

\def\lsim{\mathrel{
\rlap{\raise 0.511ex \hbox{$<$}}{\lower 0.511ex
\hbox{$\sim$}}}}

\begin{document}

\begin{center}
{\large{\bf COSMIC STRINGS AND THE STRING DILATON}}
\end{center}
\vskip .1in
\begin{center}
{Thibault Damour}

{\it Institut des Hautes Etudes Scientifiques}

{\it F-91440, Bures sur Yvette, France, and}

{\it DARC, CNRS-Observatoire de Paris,}

{\it F-92195 Meudon, France}

\medskip

{Alexander Vilenkin}

{\it Institute of Cosmology,} 

{\it Department of Physics and Astronomy}

{\it Tufts University, Medford, MA 02155, USA}

\vskip .1in

\end{center}
\vskip .5in

\begin{abstract}
The existence of a dilaton (or moduli) with
gravitational-strength coupling to matter imposes stringent constraints on
the allowed energy scale of cosmic strings, $\eta$. In particular,
superheavy gauge strings with $\eta \sim 10^{16} \, {\rm GeV}$ are ruled out
unless the dilaton mass $m_{\phi} \gsim 100 {\rm TeV}$, while the
currently popular value $m_{\phi} \sim 1 {\rm TeV}$ imposes the bound $\eta
\lsim 3 \times 10^{11} \, {\rm GeV}$.  Similar constraints are
obtained for global topological defects.  Some non-standard
cosmological scenarios which can avoid these constraints are pointed out.
\end{abstract}

\baselineskip = 1.67\baselineskip

\vskip .5in
Superstring theory predicts the existence of light gauge-neutral scalar
fields (the dilaton and the moduli) with gravitational-strength couplings to
ordinary matter. Of particular interest among those fields is the
model-independent dilaton, whose tree-level couplings are well understood. 
Because of their weak couplings, the lifetimes of the moduli can be very
long.  In a cosmological context, if moduli are created in the early
universe, their slow decay rate is the source of serious potential conflicts
with observations \cite{CFKRR}, \cite{ENQ}, \cite{CCQR}, \cite{1}.  To
simplify the discussion, we shall refer to moduli as ``the dilaton'', but
most of the following treatment is applicable, {\it mutatis mutandis}, to
generic moduli.

Several mechanisms of cosmological dilaton production have been discussed in
the literature.  First, the value of the dilaton field in the early
universe can be set away from the minimum of its potential \cite{CFKRR},
\cite{ENQ}, \cite{CCQR}, \cite{1}.   (This is
usually the case because the minima of the dilaton effective potential
at late and early times generically differ by $O (m_{\rm Planck})$). 
Coherent oscillations of the field about the minimum are then equivalent to a
condensate of nonrelativistic particles.  Another mechanism is the
production of dilatons in binary particle collisions in a hot plasma 
\cite{2}.
A third production mechanism is the amplification of quantum fluctuations of
the dilaton field in early cosmology \cite{G94}, \cite{DV96}. Requiring
consistency between the cosmological production of dilatons and observations
leads to very stringent, and {\it a priori} unnatural, constraints on the
dilaton mass and couplings \cite{ENQ}. Several mechanisms have been proposed
to solve this cosmological moduli problem: e.g. a late stage of secondary
inflation \cite{2}, \cite{18}, or the presence of a symmetry of moduli space
ensuring the coincidence of the minima of the effective potential at early
and late times \cite{19}, \cite{DV96}.

In this paper we shall discuss another mechanism of dilaton
production.  Oscillating loops of cosmic string, which could be formed
at a symmetry-breaking 
phase transition in the early universe, will copiously emit dilatons,
as long as the characteristic frequency of oscillation is greater than
the dilaton mass. Cosmic strings are predicted in a wide class of elementary
particle models \cite{4}.  Their mass per unit length $\mu$, which is
equal to the string tension, is determined by the symmetry breaking
energy scale $\eta$, $\mu\sim\eta^2$ \cite{5}.  Of particular interest
are grand-unification strings with $\eta\sim 10^{16}~{\rm GeV}$ which could
be responsible for the formation of galaxies and large-scale
structure.  We shall calculate the dilaton density produced by the
strings and
explore the constraints it imposes on the dilaton and string
parameters.  

We assume that  the string thickness is small compared to the
loop size and to the Compton wavelength of the dilaton, so that the
string can be regarded as an infinitely thin line. We work in the ``Einstein
conformal frame'' where tensor gravity decouples from the dilaton and is
described by the standard Einstein-Hilbert action. Then the interaction of the
dilaton field $\phi$ with the string is described by the action 
\begin{equation} 
S=-{1\over{4\pi G}}\int
d^4x\left[{1\over{2}}(\nabla\phi)^2 +V(\phi) \right] -\int\mu (\phi)dS.
\label{1}
\end{equation}
Here, $\mu(\phi)$ is the $\phi$-dependent string tension, $dS$ is the
surface element on the string worldsheet, $G$ is Newton's
constant, and we have used the Minkowski metric assuming the spacetime
to be approximately flat.

We choose the origin of $\phi$ so that the minimum of the
dilaton potential $V(\phi)$ is at $\phi=0$.  Then, for $\phi$ in the
vicinity of the minimum, the dilaton field equation takes the form
\begin{equation}
(\nabla^2 -m_\phi^2)\phi(x)=-4\pi G\alpha T(x),
\label{2}
\end{equation}
where $m_\phi =[V''(0)]^{1/2}$ is the dilaton mass,
and $T(x)$ is the trace of the string
energy-momentum tensor.  The dimensionless parameter $\alpha \equiv \partial
{\rm ln} \sqrt{\mu (\phi)} / \partial \phi =\mu'(0)/2\mu(0)$ measures the
strength of the coupling of $\phi$ to cosmic strings. One generically expects
$\alpha\sim 1$.  

The string world history can be represented as $x^\mu (\zeta^a)$,
$a=0,1$, where $\zeta^0$ and $\zeta^1$ are the worldsheet coordinates.
The choice of these coordinates is largely arbitrary; it is convenient for
most purposes to use a ``conformal gauge'', specified by the conditions

\begin{equation}
{\dot x}\cdot x'=0, ~~~~~~~~~~~~ {\dot x}^2+{x'}^2=0,
\label{3}
\end{equation}
where dots and primes stand for differentiation with respect to
$\zeta^0$ and $\zeta^1$, respectively.  The residual freedom of
coordinate transformations can be used to set $\zeta^0=x^0\equiv t$,
which allows us to describe the string trajectory using the
three-vector ${\bf x}(\zeta, t)$, where $\zeta\equiv \zeta^1$.  
The string energy-momentum tensor is then given by
\begin{equation}
T^{\mu\nu}(x,t)=\mu\int d\zeta ({\dot x}^\mu {\dot x}^\nu
-{x'}^\mu{x'}^\nu) \delta^{(3)} 
[{\bf x}-{\bf x}(\zeta,t)],
\end{equation}
and its trace is
\begin{equation}
T(x,t)=-2\mu\int d\zeta {\bf x'}^2(\zeta,t)\delta^{(3)} 
[{\bf x}-{\bf x}(\zeta,t)],
\label{5}
\end{equation}
where $\mu\equiv\mu (\phi =0)$.   Disregarding dilatonic and
gravitational back-reaction, the string equation of motion in the
gauge (\ref{3}) has a simple form,
\begin{equation}
{\ddot {\bf x}}-{\bf x''}=0.
\label{6}
\end{equation}
It can be shown from Eqs. (\ref{3}), (\ref{6}) that the motion of a
closed loop of string is periodic with a period $L/2$, where $L\equiv
M/\mu$ and $M$ is the loop's mass.  The quantity $L$ is often called
the length of the loop, although the actual length varies with time.

The rates of dilaton energy loss and of dilaton number production by a
periodic source of angular frequency $\omega$ can be found from the following
general equations
\begin{equation}
{\dot E}_{\phi}=\sum_n P_n,~~~~~~~~~~~~~~~~~ {\dot N}_{\phi}=\sum_n
P_n/\omega_n, \label{7}
\end{equation}
\begin{equation}
P_n ={G\alpha^2\over{2\pi}}\omega_n k_n \int d\Omega 
|T({\bf k},\omega_n)|^2 ,
\label{8}
\end{equation}
\begin{equation}
T({\bf k},\omega_n)={1\over{T_n}}\int_0^{T_n} dt\int d^3 x
e^{i\omega_n t-i{\bf kx}}T({\bf x},t),
\label{9}
\end{equation}
where $\omega_n=n\omega$, $T_n=2\pi/\omega_n$, $n=1,2, ...$; $k_n \equiv
|{\bf k}|=(\omega_n^2 -m_\phi^2)^{1/2}$, $d\Omega$ is the solid angle
element, and the angular integration is over the directions of ${\bf
k}$.  The dilaton momentum $k_n$ has to be real; hence, only terms with
$\omega_n>m_\phi$ are included in the sums (\ref{7}).

For a loop of length $L$, $\omega_n=4\pi n/L$, and the sums are taken
over $n>L/L_c$, where
\begin{equation}
L_c=4\pi/m_\phi.
\label{10}
\end{equation}
For $L \ll L_c$, $\omega_n \gg m_\phi$ for all values of $n$, and we
can approximately set $m_\phi=0$.  Then, dilaton radiation from
specific loop trajectories (described by solutions of Eqs. (\ref{3}),
(\ref{6})) can be analyzed using the techniques developed for the
gravitational case in Ref. \cite{6}.  Details of such an analysis will
be given in a separate paper \cite{7}; here we shall only summarize
the results.  We found that the energy spectrum and angular
distribution of the dilaton radiation are very similar to the
gravitational case (and very different from the electromagnetic
radiation by superconducting strings \cite{8}).  The energy and
particle radiation rates can be represented as
\begin{equation}
{\dot E}_\phi=\Gamma_\phi \alpha^2 G\mu^2,~~~~~~~~~~~~~ {\dot N}_\phi ={\tilde
\Gamma}_\phi \alpha^2 G\mu^2/\omega,
\label{11}
\end{equation}
where the numerical coefficients $\Gamma_\phi$ and ${\tilde \Gamma}_\phi$
depend on the loop trajectory (but not on its size).  Typically, $\Gamma_\phi
\sim 30$ and ${\tilde \Gamma}_\phi \sim 13$. The total radiation power
from the loop is ${\dot E}=\Gamma G\mu^2$ with $\Gamma=\Gamma_g
+\alpha^2 \Gamma_\phi$, where $\Gamma_g \sim 65$.  The high-frequency
asymptotic of the spectrum for a loop with cusps is $P_n\propto n^{-4/3}$,
and for a cuspless loop with kinks is $P_n\propto n^{-2}$.  This can be used
to estimate the radiation rates from loops with $L\gg L_c$,
\begin{equation}
{\dot E}_\phi\sim\Gamma_\phi \alpha^2 G\mu^2(L/L_c)^{-1/3},~~~~~~~~~ {\dot
N}_\phi \sim{\tilde \Gamma}_\phi \alpha^2 G\mu^2 m_\phi^{-1}
(L/L_c)^{-1/3}. 
\label{12}
\end{equation}
Here, we used a `cuspy' loop spectrum, $P_n\propto n^{-4/3}$, and
introduced a lower cutoff at $n\sim L/L_c$.

To estimate the cosmological density of dilatons produced by
oscillating string loops, we shall adopt a simple model in which the
loops radiate all $\phi$-quanta in the fundamental mode, $\omega_1
\equiv\omega =4\pi/L$.  This approximation has been proven to give
accurate results (within a factor of $\sim 3$) for the gravitational
wave power spectrum.  Moreover, the large-$n$ contribution to ${\dot
N}_\phi$ in Eq. (\ref{7}) converges faster than that to the power
${\dot E}_\phi$, and thus we expect our estimate for the particle
density to be no less accurate than that for the power spectrum.

Loops of initial length $L$ are chopped off the string network at a
cosmic time $t_i\sim L/\beta$ \cite{N} and decay at time $t_f\sim(\Gamma
G\mu)^{-1}L$.  They have number density $n_i\sim\zeta\beta^2/L^3$ at
the time of birth and 
\begin{equation}
n_f\sim\left({t_i\over{t_f}}\right)^{3/2} n_i\sim{\kappa^{1/2}\zeta
\over{\Gamma G\mu t_f^3}}
\label{13}
\end{equation}
at the time of decay. Here, $\zeta$ is a parameter characterizing the
density of long strings (its definition can be found in
Ref. \cite{4}), $\kappa\equiv\beta/\Gamma G\mu$, and we have used the
radiation era expansion law, $a(t)\propto t^{1/2}$.  Numerical
simulations of string evolution indicate that $\zeta\sim 14$ and
$\beta \lsim 10^{-3}$.  The exact value of $\beta$
is not known, but it is bounded from below by $\beta \gsim \Gamma G\mu$, so
that $\kappa \gsim 1$. From Eq. (\ref{11}), the total
number of dilatons emitted by a loop decaying at $t\sim t_f$ is
\begin{equation}
N\sim{\dot N}t_f \sim(4\pi)^{-1}\Gamma{\tilde \Gamma}_\phi \alpha^2 G^2\mu^3
 t_f^2.
\label{14}
\end{equation}

The quantity of interest to us will be $Y_\phi=n_\phi(t)/s(t)$.  Here,
$n_\phi(t)$ is the dilaton density, $s(t)$ is the entropy density,
which during the radiation era is given by $s(t)=0.0725 [{\cal
N}(t)]^{1/4} (m_p/t)^{3/2}$, ${\cal N}(t)$ is the effective number of
spin degrees of freedom at time $t$, and $m_p$ is the Planck mass.
Apart from dilaton production and decay and out-of-equilibrium phase
transitions (such as thermalization after inflation), $Y_\phi$ is
conserved in the course of the cosmological evolution.  The
contribution to $Y_\phi$ from loops decaying at $t\sim t_f$ can be
estimated as
\begin{equation}
Y_\phi (t_f)\sim n_f N/s(t_f)\sim\kappa^{1/2}\zeta{\tilde \Gamma}_\phi
\alpha^2(G\mu)^2 (m_p t_f)^{1/2}{\cal N}_f^{-1/4},
\label{15}
\end{equation}
where ${\cal N}_f\equiv {\cal N}(t_f)$.  Eqs. (\ref{14}) and
(\ref{15}) are valid as long as $t_f \lsim t_c\equiv
4\pi/\Gamma G\mu m_\phi$, so that the loop sizes are smaller than the
critical size (\ref{10}).  From Eq. (\ref{15}) we see that larger
values of $t_f$ give a greater contribution, and thus the dominant
contribution to $Y_\phi$ is given by $t_f\sim t_c$ \cite{9}.  With
$\zeta\sim 14$, ${\cal N}(t_c)\sim 100$, ${\tilde \Gamma}_\phi \sim
13$ and $\Gamma\sim 100$, we have
\begin{equation}
Y_\phi\sim Y_\phi(t_c)\sim 20\kappa^{1/2}\alpha^2(G\mu)^{3/2}
(m_p/m_\phi)^{1/2}. 
\label{16}
\end{equation}
Eq. (\ref{16}) is the main result of the present paper.

Strings of energy scale $\eta$ are typically formed at time $t_s\sim
t_p/G\mu$, where $t_p=m_p^{-1}$is the Planck time.  Long strings 
are initially
overdamped and begin to move relativistically at time $t_*\sim
t_p/(G\mu)^2$.  Small loops become relativistic at an earlier time,
but damping due to interaction with the surrounding plasma remains a
significant energy loss mechanism until $t_*$.  
In the derivation of Eqs. (\ref{11}), (\ref{16}) we 
assumed damping to be negligible, and thus the condition of validity
of (\ref{16}) is $t_c >t_*$, which gives 
\begin{equation}
m_\phi/m_p < 4\pi G\mu/\Gamma.
\label{16*}
\end{equation}

The analysis of the cosmological implications of the dilaton density
(\ref{16}) is similar to that for any weakly-interacting relic
particles \cite{ENQ}, \cite{3}.  The resulting constraints are sensitive to
the lifetime of the dilaton $\tau$, which is determined by its mass and
couplings.  The dilaton couples (in the Einstein frame) to spin-0 and spin-1/2
particles only through the mass terms, so that decays into such particles are
suppressed by powers of their mass \cite{10}.  The interaction
Lagrangian responsible for decays into light gauge bosons is ${\cal
L}_{\rm int} = \frac{1}{2}\alpha_F\phi F_{\mu\nu}^2$, and the corresponding
lifetime is
\begin{equation}
\tau=4m_p^2/N_F\alpha_F^2m_\phi^3=3.3\times 
10^{13}(12/N_F)\alpha_F^{-2} m_G^{-3} ~s.
\label{17}
\end{equation}
Here, $m_G=m_\phi/1~{\rm GeV}$, $N_F$ is the number of gauge bosons with
masses $\ll m_\phi$, and the value of $\alpha_F^2$ is averaged over
all such bosons.  For $m_\phi \gsim 1~{\rm TeV}$, all
standard-model gauge bosons should be included ($N_F=12$).  The coupling
constant $\alpha_F$ is normalized so that $\alpha_F =1$ for a tree-level
superstring dilaton. It is generically expected that $\alpha_F \sim 1$ for
all moduli.  For numerical estimates below we set $\alpha_F=\alpha=\kappa=1$. 
(Note that since $Y_\phi\propto \kappa^{1/2}$ and $\kappa\gsim1$, setting
$\kappa=1$ will result in conservative bounds on $\mu$ and $m_\phi$).

A multitude of astrophysical constraints on unstable relic particles
have been discussed in the literature. Short-range Cavendish experiments
\cite{MP88} exclude ultra-light dilatons of mass smaller than $1.6 \times
10^{-3} {\rm eV}$ \cite{20}. For quasi-stable dilatons, with lifetimes
larger than the present age of the universe, $\tau > t_0 \simeq 4 \times
10^{17} s$ (corresponding to $m_{\phi} \lsim 40 {\rm MeV}$), one has the
usual upper bound on the cosmological dilaton mass density $\Omega_{\phi}
h^2 < 1$ \cite{KT}, where $\Omega_{\phi} = n_{\phi} m_{\phi} / \rho_{\rm
critical}$ and $h \equiv H_0 / 100 {\rm km} s^{-1} {\rm Mpc}^{-1}$. This
yields $Y_{\phi} < 3.6 \times 10^{-9} m_G^{-1}$. For $\tau\gsim
t_{\rm dec}\sim 10^{13}~s$, very stringent constraints follow from the
limits on the diffuse $\gamma$-ray background that would result from dilaton
decays \cite{KT}: $Y_\phi < 2.9\times 10^{-16} m_G^{-1}$ for
$t_{\rm dec}\lsim\tau\lsim t_0$ and $Y_\phi <1.3\times
10^{-20}m_G^{-4}$ for $\tau\gsim t_0$.  For $10^{-1}~s \lsim\tau\lsim
t_{\rm dec}$, the bounds are obtained by requiring that the decay products do
not significantly change the abundances of $^4{\rm He}$, $^3{\rm He}$ and
D.  The relevant processes are the interaction of ambient nucleons with the
hadronic showers resulting from hadronic decays for $0.1~s
\lsim\tau\lsim10^7~s$ \cite{12}, \cite{Reno}, and photodissociation and
photoproduction of light elements by electromagnetic cascades initiated by
the decay products for $10^4~s\lsim\tau\lsim t_{\rm dec}$
\cite{13,3} (both processes being important for $10^4 \, s \lsim \tau \lsim
10^7 \, s$ \cite{12}). The $\tau$-dependence of the resulting bound on the
dilaton density is rather complicated, but roughly $Y_\phi \lsim 1.4\times
10^{-12}m_G^{-1}$ for $10^7~s\lsim \tau\lsim t_{\rm dec}$ and $Y_\phi \lsim
10^{-13} - 10^{-14}$ for $1~s\lsim\tau \lsim 10^7~s$.  For $\tau < 0.1~s$,
dilatons decay well before the onset of nucleosynthesis, and the bound is
rapidly weakened as we move towards smaller values of $\tau$.  

Combining these bounds on $Y_\phi$ with the expression (\ref{16}) for
the dilaton density produced by cosmic strings, we obtain constraints
on $m_\phi$ and $\mu$ which are represented in Fig.1.  We see that the
excluded domain cuts deeply into the region of physically interesting
values of the parameters.  In particular, the most popular values of
$G\mu\sim 10^{-6}$ and $m_\phi\sim 1~{\rm TeV}$ are incompatible with one
another.  If the dilaton mass is indeed $\sim 1~{\rm TeV}$, then the string
tension is bounded by $G\mu\lsim 6 \times 10^{-16}$, which
corresponds to symmetry breaking scales $\eta\lsim
3 \times 10^{11}~{\rm GeV}$.  On the other hand, if GUT-scale strings are
discovered, then the dilaton mass must satisfy $m_\phi\gsim 100~{\rm TeV}$. 

These conclusions are rather robust with respect to the variation of
the numerical coefficient in Eq. (\ref{16}) (which we expect to be
accurate only within a factor of $\sim 3$).  If, for example, the
coefficient is changed by one order of magnitude, then the bound on
$G\mu$ at a fixed $m_\phi$ is modified by a factor of $\sim 5$, and
the bound on $m_\phi$ with $G\mu$ in the grand unification range
remains essentially unchanged.

In the derivation of Eq. (\ref{16}) for $Y_\phi$ we assumed that
gravitational and dilaton radiation were the dominant energy loss
mechanisms of strings.  This is justified for gauge strings, formed as
a result of a gauge symmetry breaking.  In the case of global strings,
oscillating loops lose most of their energy by Goldstone boson
radiation at the rate ${\dot E}\sim\Gamma\eta^2$.  Here, $\eta$ is the
global symmetry breaking scale and $\Gamma$ is about the same as in
the gravitational case, $\Gamma\sim 65$.  The mass per unit length of
a global string has a logarithmic length-dependence, $\mu=2\pi\eta^2
\ln (L/2\pi\delta)$, where $\delta\sim\eta^{-1}$ is the string
thickness, and the lifetime of a loop is $\tau\sim E/{\dot E}\sim KL$
with $K=(2\pi/\Gamma)\ln (L/2\pi\delta)$.  If we take, for example, a
GUT-scale string with $\eta\sim 10^{15}~{\rm GeV}$ of length $L\sim
L_c=4\pi/m_\phi$ with $m_\phi\sim 1~{\rm TeV}$, then $K\sim 3$, and a loop
will make $\sim 6$ oscillations before losing half of its energy.  The
loops are expected to form with sizes $L\sim t/K$ and decay in about a
Hubble time: $t_i\sim t_f\sim KL$.  Once again, the main contribution
to $Y_\phi$ comes from $t_f\sim t_c\sim 4\pi K/m_\phi$, and it is
easily verified that Eq. (\ref{16}) is replaced by
\begin{equation}
Y_\phi \sim 350\alpha^2(G\mu)^2(m_p/m_\phi)^{1/2},
\label{18}
\end{equation}
and its condition of validity (\ref{16*}) by
\begin{equation}
m_\phi/m_p <40(G\mu)^2,
\label{19}
\end{equation}
where we have used $K\sim 3$.  For $m_\phi\sim
1~{\rm TeV}$ the bound on the dilaton density is $Y_\phi\lsim 3\times
10^{-14}$, and the constraint on $G\mu$ following from Eq. (\ref{18}) (with
$\alpha \sim 1$) is  $G\mu \lsim 10^{-12}$.  However, according to
(\ref{19}), with this value of $m_\phi$ Eq. (\ref{18}) is valid only
for $G\mu \gsim 2\times 10^{-9}$, and thus we can
conclude only that $G\mu\lsim 2\times 10^{-9}$.

Global monopoles and textures have also been suggested as possible
seeds of galaxy formation \cite{14,15}.  The energy density of these
defects varies on the horizon scale $R\sim t$.  The corresponding
field gradients are ${\dot \Phi}\sim|{\bf \nabla}\Phi|\sim\eta/t$ and
$T_\mu^\nu\sim\eta^2/t^2$.  The dilaton density produced by the
relativistic evolution of the field $\Phi$ in a Hubble time can be
estimated from Eq. (\ref{2}): $\phi/t^2\sim 4\pi G\alpha\eta^2/t^2$,
which gives $\phi\sim 4\pi G\alpha\eta^2$ and $n_\phi\sim m_\phi\phi^2
/8\pi G$.  With $s\sim 0.2(m_p/t)^{3/2}$ and $t\sim t_c\sim
4\pi/m_\phi$, we have
\begin{equation}
Y_\phi\sim 10^3\alpha^2 (\eta/m_p)^4(m_p/m_\phi)^{1/2}.
\label{20}
\end{equation}
With $\alpha=1$ and $m_\phi \sim 1~{\rm TeV}$, the resulting constraint on
$\eta$ is $\eta\lsim 10^{13}~{\rm GeV}$.  [Damping is
unimportant for global monopoles and textures \cite{damping}, 
and there is no
analogue of the condition (\ref{19})].  Hence, gauge
cosmic strings and global strings, monopoles and textures are excluded
as seeds for structure formation if the dilaton mass is $m_\phi\sim
1~{\rm TeV}$.

We finally mention some ways of avoiding the above constraints.  The
main contribution to the dilaton density in Eqs. (\ref{16}),
(\ref{18}) and (\ref{20}) comes from the time $t\sim t_c$, which
corresponds to the temperature $T_c\sim 10^9(G\mu)^{1/2}m_G^{1/2}~{\rm GeV}$
for gauge strings and $T_c\sim 10^8m_G^{1/2}~{\rm GeV}$ for global defects.
Our analysis, therefore, is not directly applicable to models in which
the universe has never been heated up to such temperatures.  For
example, in inflationary scenarios the thermalization temperature
after inflation can be below $T_c$.  Alternatively, string formation
can be delayed until after $T_c$: in some supersymmetric models
GUT-scale strings can be formed at temperatures as low as the
electroweak scale \cite{16}.  Another possibility is to invoke models
where topological defects are produced during inflation \cite{17}.
Then the defects begin emitting dilatons only after their
characteristic scale comes within the horizon, which can happen at
$t>t_c$.  In all three cases the resulting dilaton density is very
model-dependent.

Once dilatons are produced, they can be diluted by a brief period of
inflation.  Models of this kind have been suggested \cite{2}, \cite {18} to
resolve the usual Polonyi-moduli problem: overproduction of dilatons
and other moduli due to a mismatch of the minima of their potential at early
and late times.  The same models can be used to relax the
constraints on topological defects discussed here.  We note, however, that
another proposed solution to the moduli problem will not work in our case. 
Dine, Randall and Thomas \cite{19} suggested that moduli production during
inflation can be suppressed if the potential in moduli space has some
symmetry which enforces that the potential minima before and after inflation
coincide.  Clearly, this does not resolve the conflict between moduli
and defects: all defects formed after inflation will produce dilatons,
and thus the defect parameters are subject to all constraints we
discussed earlier in this paper.  The only exception is the model
suggested in Ref. \cite{20} (whose cosmological consequences were further
studied in Ref. \cite{DV96}) in which the minimum of the potential is a point
of enhanced symmetry for all dilaton couplings.  Then, near the minimum, the
dilaton is essentially decoupled from all other fields (in particular
$\alpha \ll 1$), and dilaton production by topological defects is suppressed.

\bigskip

A.V. is grateful to IHES (France), where most of this work was done, 
for hospitality, and to the National Science Foundation for partial
support.

\begin{figure}[h]
\caption{\label{FiG.1.}{Constraints on $\log_{10} (G\mu)$ versus $\log_{10}
(m_{\phi} / 1 {\rm Gev})$. The region above the solid curve is forbidden.
Labels indicate the source of the various constraints: Cavendish experiments
(C), $\Omega_{\phi} h^2 < 1(O)$, gamma-ray background (G), photodissociation
(P), combined hadroproduction and photodissociation (HP), hadroproduction (H).
The dashed line indicates the condition of validity (17). [The constraints
apply only above the dashed line.]}}
\end{figure}


\begin{thebibliography}{11111}
\bibitem{5}We use units in which $\hbar =c=1$ and the metric
signature $-+++$.
\bibitem{N}The dimensionless parameter $\beta$ describing the scaling of
string loops is usually denoted $\alpha$.
\bibitem{9}It is easily verified using Eq. (\ref{12}) that loops with
$L>L_c$ give a subdominant contribution.
\bibitem{10}The only exception is the axion partner, $a$, of the dilaton
$\phi$ which also has a kinetic-term coupling, $\propto e^{4\alpha_a \phi}
(\nabla a)^2$, with $\alpha_a =1$ for a tree-level coupled superstring
dilaton.  If the axion mass is $m_a \ll m_\phi$, then the factor
$N_F\alpha_F^2$ in Eq. (\ref{17}) should be replaced by $N_F\alpha_F^2
+2\alpha_a^2$.
\bibitem{damping}The damping force on a global monopole moving with
velocity $v$ through a relativistic plasma of temperature $T$ can be
estimated as $F_d\sim n\sigma v_T\Delta p$, where $n\sim T^3$ is the
particle density, $\sigma\sim T^{-2}$ is the particle scattering cross
section off the monopole, $v_T\sim 1$ is the thermal velocity, and
$\Delta p \sim Tv$ is the average momentum transfer per collision.
This gives $F\sim T^2 v$.  On the other hand, the interaction force
between monopoles is $F\sim 4\pi\eta^2$ and is always much greater
than $F_d$, except perhaps immediately after the monopole-forming
phase transition ($T\sim \eta$).
\end{thebibliography}
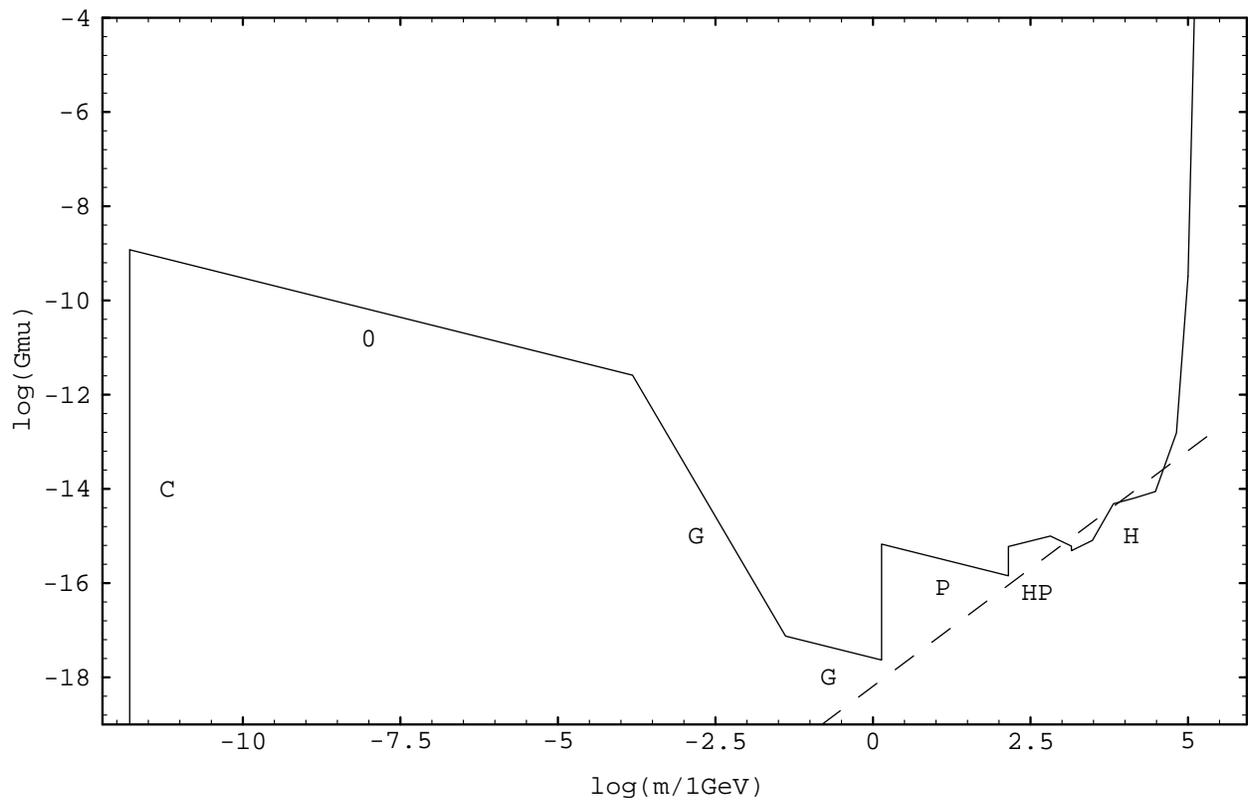
\end{document}